\documentclass[amsart]{IEEEtran} 
\usepackage{amsmath,amssymb,mathtools}
\usepackage{balance}
\usepackage{color}
\usepackage{graphicx,bm}
\usepackage{caption}
\usepackage{subcaption}
\usepackage{multicol}
\usepackage{csquotes}
\usepackage[T1]{fontenc}
\usepackage{cite}
\usepackage{epstopdf}
\usepackage{framed}
\usepackage{hyperref}
\usepackage{url}
\usepackage{multirow}
\usepackage{fancyhdr}
\begin{document}
\lhead{This work has been submitted to the IEEE Wireless communications letters for possible publication. Copyright may be transferred without notice, after which this version may no longer be accessible.}
\title{PCI-MDR: Missing Data Recovery in Wireless Sensor Networks using Partial Canonical Identity Matrix}

\author{Neha Jain \textsuperscript{$\ast$}\textsuperscript{$\dagger$}, Anubha Gupta\textsuperscript{$\dagger$}, Vivek Ashok Bohara\textsuperscript{$\ast$}
\thanks{The authors are with the Department of Electronics and Communication,
\textsuperscript{$\ast$}Wirocomm Research Group, \textsuperscript{$\dagger$}Signal processing and Bio-medical Imaging Lab, IIIT-Delhi, New Delhi, India, 110020 (e-mail: nehaj@iiitd.ac.in, anubha@iiitd.ac.in, vivek.b@iiitd.ac.in).}}

\maketitle
\thispagestyle{fancy}

\begin{abstract}
Data loss in wireless sensor networks (WSNs) is quite prevalent. Since sensor nodes are employed for various critical applications, accurate recovery of missing data is important. Researchers have exploited different characteristics of WSN data, such as low rank, spatial and temporal correlation for missing data recovery. However, the performance of existing methods is dependent on various factors. For instance, correct rank estimation is required for exploiting the low-rank behaviour of WSNs, whereas correlation information among the nodes should be known for exploiting spatial correlation. Further, the amount of missing data should not be massive for exploiting temporal correlation. To overcome the above-mentioned drawbacks, a novel method `\textit{PCI-MDR}' has been proposed in this paper. It utilizes compressive sensing with partial canonical identity matrix for the recovery of missing data in WSNs. To validate the proposed method, the results have been obtained on the real dataset of temperature sensors from the Intel Lab. The proposed method is observed to perform superior to the existing methods, yielding significant improvement.

\end{abstract}

\IEEEpeerreviewmaketitle
\section{Introduction}
In recent years, with the advent of smart sensors, the design and deployment of wireless sensor networks (WSNs) have become an active area of research. Sensor nodes are typically deployed to measure environmental parameters/conditions such as temperature, pressure, humidity, and vibration, etc. The measured value is transmitted to the fusion centre (FC) for further processing. In WSN applications, data loss is prevalent and generally arises due to hardware failures, channel fading, synchronization issues, collisions, and environmental blockages \cite{4650774}. These missing values are required to be estimated accurately. The incorrect estimation of the missed raw data can lead to serious damage or casualty. For example, the underwater temperature measurements are used to determine the nature of ocean currents, helping in generating environmental alerts in case of any adversity. The above discussion clearly establishes the need for designing effective methods for recovering missing data in WSNs. 

A number of different methods have been employed to estimate missing values in the context of WSNs \cite{neter1996applied,cover1967nearest,pan2010k,gao2015missing,6566962}. Linear interpolation (LA) is one of the simplest methods that interpolates the data in the time domain to determine the missed values. However, the performance of LA is unsatisfactory if there is continuous data loss or data is collected at long intervals. Linear regression \cite{neter1996applied} and \textit{K}-nearest neighbours (KNN) \cite{cover1967nearest} exploit the spatial characteristics of the WSN for missing data recovery. In the KNN method, the accuracy of the estimated values is highly dependent on the number of  \textit{K} neighbours. In \cite{pan2010k}, linear regression and KNN are exploited jointly for data recovery. In this, parameters of linear models, corresponding to each pair of nodes, are learned. If data is missed from any one node, values are estimated by the remaining nodes using the learned model. On the basis of spatial correlation among the nodes, it assigns weights to each estimated value for the data recovery. The above methods are effective when the correlation coefficients between the nodes are known. Unlike the above methods, \cite{gao2015missing} exploited both the time and spatial correlation. In this method, a weighted combination of spatial and temporal estimation is considered. However, all the above methods require the correlation information, which would not be known precisely for incomplete data matrix. Similarly, the low-rank behaviour of WSN data along with the spatial and temporal profiles is exploited in \cite{6566962}. However, this method requires the accurate rank of the WSN data matrix. Further, real WSN dataset is, in general, not exactly low-rank, but are usually approximated as low-rank. Moreover, in order to exploit spatial correlation, the algorithm requires the topology of the network, which is generally not available as sensor nodes are randomly placed. Hence, good performance of the algorithms in \cite{6566962} can be attained only when the above conditions are satisfied. 

In this paper, \textit{Partial canonical identity matrix based missing data recovery (PCI-MDR)} algorithm has been proposed for WSNs. The proposed method overcomes the above-mentioned drawbacks of the existing methods. Some of the major contributions of the proposed work are: 
\begin{itemize}
\item The problem of missing data has been formulated in terms of compressive sensing (CS) framework through the following steps. 
\begin{itemize}
\item[1.] The position of missing data is captured via constructing the corresponding PCI sensing matrix.
\item[2.] Then, we have utilized the work in \cite{100000}, which shows that DCT acts like a Karhunen-Loeve (KL) type basis for a large class of smooth signals, hence may act as one of the best sparsifying basis for smooth signals in CS based reconstruction.
\item[3.] Leveraging the insights from the above, we have evaluated the coherency of PCI and DCT matrices and have shown that they are highly incoherent. Hence, this combination is appropriate for CS based reconstruction.
\item[4.] After recovering the data using PCI and DCT based CS framework, a de-noising step has been used at the second stage along with low-rank constraint to further improve the performance of data recovery.
\end{itemize}

\item In literature, the problem of matrix completion is solved by exploiting low-rank behaviour of the data using algorithms such as matrix factorization \cite{5961765} and nuclear norm minimization \cite{cai2010singular}. In this work, we have shown that even if the data is low-rank, the proposed PCI-MDR method outperforms various matrix completion algorithms.

\item In the case of noisy WSN data, the second stage of de-noising in the proposed method helps in further improving the performance of data recovery. 

\item As shown later, the proposed method can be applied in realistic scenarios as no prior information such as rank and topology of the data is required.

\item The accuracy of the PCI-MDR method is tested on the real dataset of temperature, taken from Intel Lab \cite{123123}. Simulation results demonstrate the superior performance of PCI-MDR compared to the conventional matrix completion algorithms.
\end{itemize}
 
\textit{Notations}: Transpose of a matrix or vector is denoted as $(\cdot)^T$. Matrices are represented in capital and bold letters, vectors in small-case and bold letters, and variables are written in italics. `$\bullet$' denotes element wise multiplication and $||.||_p$ represents the $l_p$ norm. If $\textbf{x}$ is the original signal vector, $\hat{\textbf{x}}$ represents the recovered signal. 
\section{Background}
Consider a WSN matrix $\textbf{X}_{n\times t}$, where $n$ is the number of sensor nodes and $t$ is the number of time points. The received incomplete matrix can be written as $\textbf{Y}=\textbf{B} \bullet \textbf{X} $, where `$\bullet$' denotes element wise multiplication and \textbf{B} is a binary index matrix such that $B(i,j)=1$, if $X(i,j)$ is present and is 0 if $X(i,j)$ is missing. If $\textbf{X}$ is low-rank, it can be recovered by solving the following convex optimization program \cite{candes2009exact}:
\begin{equation}
\min_{\textbf{X}} \ \  \text{rank}(\textbf{X})\ \ \ \text{s.t}  \ \ \ \ \ \textbf{Y}=\textbf{B} \bullet \textbf{X}. \label{1}
\end{equation}
However, this is an NP-hard problem. Moreover, real dataset are not necessarily \textit{exactly low-rank}, instead these may be \textit{approximately low-rank}. A signal is called \textit{exactly low-rank} if it has only a few number of non-zero singular values and rest are zero. On the other hand, a signal is \textit{approximately low-rank} if it has a few singular values with large coefficients and remaining values tending to zero. Hence, instead of minimizing the rank of \textbf{X}, we can also minimize the sum of the singular values \cite{recht2010guaranteed}. Therefore, (\ref{1}) reduces to
\begin{equation}
\min_{\textbf{X}}  ||\textbf{X}||_* \ \ \ \text{s.t.}  \ \ \ \ \ \textbf{Y}=\textbf{B} \bullet \textbf{X}. \label{2}
\end{equation}
where $||\textbf{X}||_*$ is the nuclear norm of \textbf{X}. (\ref{2}) can be solved by \textit{singular value thresholding (SVT)} algorithm \cite{cai2010singular}.
\textit{Matrix factorization} is also one of the famous techniques for matrix completion \cite{5961765}, if rank of \textbf{X} (say \textit{r}) is known. In such a scenario, (\ref{1}) can be written as:
\begin{equation}
\min_{\textbf{U},\textbf{V}} (||\textbf{U}||_F^2 + ||\textbf{V}^T||_F^2) \,\,\text{  s.t.}  \ \ \ \ \textbf{Y}=\textbf{B} \bullet (\textbf{UV}^T). \label{3}
\end{equation}
Here, we need to find matrices $\textbf{U}$ and $\textbf{V}$ such that $\textbf{X}=\textbf{UV}^T$. The dimensions of $\textbf{U}$ and $\textbf{V}$ are $N \times r$ and $T \times r$, respectively. In \cite{6566962}, authors exploited three factors: rank, spatial and temporal correlation, for better data recovery by solving:
\begin{align}
\min_{\textbf{U},\textbf{V}}&(||\textbf{B}\bullet(\textbf{UV}^T)-\textbf{Y}||_F^2+ \lambda (||\textbf{U}||_F^2+||\textbf{V}^T||_F^2)\nonumber\\
&+||\textbf{HUV}^T||_F^2+||\textbf{UV}^T\textbf{T}||_F^2),
\end{align}
where $\lambda$ is the Lagrange multiplier. $\textbf{T}$ and $\textbf{H}$ matrices are used to exploit the temporal and spatial correlation, respectively. Therefore this method requires both rank and topology of the data for correct estimation of missing values, which is not known in practical scenario for incomplete data matrix.

\section{Proposed \textit{PCI-MDR} Method}
In this section, we present the proposed PCI-MDR method for recovering the missing data in WSN networks. According to the theory of compressive sensing (CS) \cite{foucart2013mathematical}, the original data \textbf{x} of length $N$ can be compressively sensed by using a sensing matrix $\bm{\Phi}$ of size $M \times N$ ($M<N$) such that $\textbf{y}=\bm{\Phi}\textbf{x}$. If data $\textbf{x}$ is sparse in some domain, say $\bm{\Psi}$, then $\textbf{x}$ can be reconstructed from $\textbf{y}$ if $\bm{\Psi}^{-1}$ and $\bm{\Phi}$ are incoherent. 
Thus, the central idea in CS is to sense fewer data samples and reconstruct the full signal. This motivates us to explore \textit{CS directly} for missing data recovery in WSNs. 

Conventionally, random Gaussian and Bernoulli matrices are used as sensing matrices in CS based reconstruction, say, for example in single pixel camera. These matrices collect \textit{M} (<\textit{N}) number of linear combinations of all data samples. However, missing data implies that it is not possible to collect linear combinations of data in the domain where the data is missing. Thus, in general, matrix completion is used to solve missing data recovery problem via matrix factorization and SVT method, as discussed in Section II.

\begin{figure}[!ht]
 \centering
 \includegraphics[scale=0.6, trim=0 10 0 20]{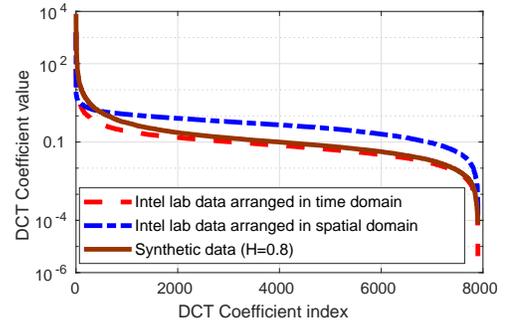}
 \caption{Sorted DCT coefficients of WSN Intel lab data \cite{123123} and synthetic data.}
 \label{fig:1}
 \vspace*{-0.3cm}
 \end{figure}

WSN data matrix $\textbf{X}_{n\times t}$, consisting of any parameter such as temperature, voltage, humidity, or light, can be vectorized either in spatial or time domain as \textbf{x} = [\textit{X}(1,1) \textit{X}(2,1).... \textit{X}(\textit{n},1) \textit{X}(\textit{n},2) \textit{X}(\textit{n}-1,2)..... \textit{X}(1,2), \textit{X}(1,3).... \textit{X}(\textit{n},\textit{t})]$^T$ and \textbf{x} = [\textit{X}(1,1) \textit{X}(1,2).... \textit{X}(1,\textit{t}) \textit{X}(2,\textit{t}) \textit{X}(2,\textit{t}-1)... \textit{X}(2,1), \textit{X}(3,1).....\textit{X}(\textit{n},\textit{t})]$^T$, respectively such that $\textbf{x}$ will be the vector of length $N=nt$. The received vector \textbf{y} of length $M$ can be written as
\begin{equation}
\textbf{y}=\bm{\Phi} \textbf{x},\label{8}
\end{equation}
where $\bm{\Phi}$ is the matrix of size $M \times N$ with $M$ as the number of available/not missed entries of full data \textbf{x}. Here, each row of $\bm{\Phi}$ contains single `1' at the position corresponding to the data samples available in $\textbf{x}$. This particular non-square matrix $\bm{\Phi}$ with a single `1' in every column and row is called the partial canonical identity (PCI) matrix.
In CS based reconstruction, \eqref{8} is written as:
\begin{align}
\textbf{y}&=\bm{\Phi} \bm{\Psi}^{-1} \textbf{s} \nonumber\\
&=\textbf{As} \ \ \ \ \ \ \ \ (\because \textbf{A}=\bm{\Phi\Psi}^{-1}),
\end{align}
where $\bm{\Psi}$ is the sparsifying transform such that data vector \textbf{x} is sparse in $\bm{\Psi}$, i.e., $\textbf{s}=\bm{\Psi}\textbf{x}$ is sparse and $\textbf{x}$ can be reconstructed from $\textbf{y}$ if $\bm{\Psi}^{-1}$ and $\bm{\Phi}$ are incoherent.
We first computed the mutual coherence, between the PCI matrix and different transform matrices used commonly in CS such as discrete cosine transform (DCT), Fourier transform (FT), and wavelet transforms (WT) in Table 1, defined as:
\begin{equation}
\mu(\mathrm{PCI},\bm{\Psi}^{-1})= \sqrt{N} \max_{\forall i,j} \frac{|\bm{\Psi}^{-1}(i,j)|}{||\bm{\Psi}^{-1}_j||_2},
\end{equation}
where $\mu(\bm{\Phi},\bm{\Psi}^{-1}) \in [1, \sqrt{N}]$ assuming $\bm{\Phi}$ and $\bm{\Psi}^{-1}$ to be $N \times N$ matrices. Here, $\mu=1$ represents maximum incoherence between the matrices.
 \begin{table}[!ht]
 \begin{center}
 \caption{Coherence between PCI matrix and transform basis}
 \begin{tabular}{|c|c||c|c|} \hline
  $\bm{\Psi}^{-1}$ & $\mu$ & $\bm{\Psi}^{-1}$ & $\mu$ \\ 
  \hline\hline
  FT or IFT & 1 & db3 wavelet & 0.8069 $\sqrt{N}$ \\ 
  \hline
  DCT or IDCT & $\sqrt{2}$ & db4 wavelet &  0.7148 $\sqrt{N}$ \\ 
  \hline
  Haar Wavelet & 0.7071$\sqrt{N}$ & Coif2 wavelet &  0.8127 $\sqrt{N}$\\
  \hline  
  db2 wavelet & 0.8365$\sqrt{N}$ & Coif4 wavelet & 0.7822 $\sqrt{N}$ \\
  \hline
  \end{tabular}
 \end{center}
 \end{table}
From Table 1, it is observed that PCI matrix is highly incoherent with Fourier transform (FT), inverse Fourier transform (IFT), discrete cosine transform (DCT) and inverse discrete cosine transform (IDCT) with coherence values of 1, 1, $\sqrt{2}$, and $\sqrt{2}$, respectively. 
Thus, if WSN data is sparse in the above-mentioned domains, \textbf{x} can be reconstructed from \textbf{y} by using the following $l^1$ minimization problem, \footnote[1]{This is a simple $l^1$ minimization that can be solved by using convex optimization methods such as SPGL1, ISTA (Iterative soft-thresholding algorithm).}
 \begin{equation}
 \hat{\textbf{x}}=\min_\textbf{x} ||\textbf{y}-\bm{\Phi}\textbf{x}||_2^2 + \lambda ||\bm{\Psi}^{-1}\textbf{x}||_1 ,\label{7}
 \end{equation}
where $\lambda$ is the regularization parameter. Indeed, the WSN data is sparse in the DCT domain as seen from the plot of sorted DCT coefficients in Fig. \ref{fig:1}. The coefficients are plotted for the temperature data taken from the Intel Lab \cite{123123}. The signal is less smooth in the spatial domain as compared to the time domain because nodes are randomly distributed in the spatial domain. Hence, from Fig. \ref{fig:1}, we observe that the data is less sparse in the spatial domain as compared to the time domain. The smoothness of the signal is captured by the parameter called Hurst exponent (H). If $H > 0.5$, the signals are smoother and their KL basis can be approximated by DCT \cite{100000}. In order to validate this, we computed the Hurst exponent, $H$ of the considered data \cite{123123} and it was observed to be nearly 0.7 in the time domain. Correspondingly, we also generated the synthetic data with $H = 0.8$. The DCT coefficients of both synthetic and real data have been plotted in Fig. \ref{fig:1}. We observe that both synthetic and real data are sparse in the DCT domain.

%
In order to exploit the sparsity due to smoothness in the spatial domain, the following problem should be solved
\begin{equation}
\min_\textbf{X} ||\textbf{Y}-\textbf{B.X}||_F^2+  \lambda_1 ||\textbf{D}^T_1\textbf{X}||_1, \label{10}
 \end{equation}
 where $\textbf{D}_1$ is the DCT matrix of size $n \times n$. Similarly, time domain smoothness can be exploited as
 \begin{equation}
\min_\textbf{X} ||\textbf{Y}-\textbf{B.X}||_F^2+  \lambda_2 ||\textbf{X}\textbf{D}^T_2||_1,\label{11}
\end{equation}
where $\textbf{D}_2$ is the DCT matrix of size  $t \times t$. These problems can be formulated using PCI sensing matrix as,
\begin{equation}
 \hat{\textbf{x}}=\min_{\textbf{x}} ||\textbf{y}-\bm{\Phi}\textbf{x}||_2^2 + \lambda_3 ||\bm{D}^{T}\textbf{x}||_1, \label{9}
 \end{equation}
where \textbf{D} is the DCT matrix of size $nt \times nt$ and \textbf{x} can be vectorized either in spatial or time domain as explained above.  To exploit double sparsity by utilizing both spatial and time domain, the following optimization problem can be solved
\begin{equation}
 \hat{\textbf{x}}=\min_{\textbf{x}} ||\textbf{y}-\bm{\Phi}\textbf{x}||_2^2 + \lambda_4 ||(\bm{D}_2 \otimes \bm{D}_1) \textbf{x}||_1. \label{12}
\end{equation}
Since, the WSN data is also low-rank, a de-noising framework with low-rank constraint is subsequently applied on the matrix recovered \footnote{$\hat{\textbf{X}}$=reshape$(\hat{\textbf{x}},[N,T])$} from (\ref{12}) as 
\begin{equation}
 \tilde{\textbf{X}}=|| \hat{\textbf{X}}-\textbf{X}||_F^2+ \lambda_5||\textbf{X}||_*, \label{13}
\end{equation}
where $\lambda_i$ (i = 1 to 5) are the regularization parameters. The matrix recovered using (\ref{12}) at the first stage and (\ref{13}) at the second stage is termed as two stage recovery process of PCI-MDR method. 

\section{Simulation Results}
In this section, we have computed the performance of the proposed PCI-MDR method for missing data recovery in both the absence as well as presence of noise. The results have been compared with matrix factorization \cite{5961765} and SVT \cite{cai2010singular} algorithms on the real data set of temperature taken from Intel Lab. In order to create the noisy data, additive Gaussian noise is added to the dataset such that the average signal-to-noise (SNR) power ratio is 10 dB. To check the robustness of the PCI-MDR, we have collected data points for a sufficiently longer time interval of one minute.
\begin{figure}[!ht]
 \centering
 \includegraphics[scale=0.5, trim=0 10 0 2]{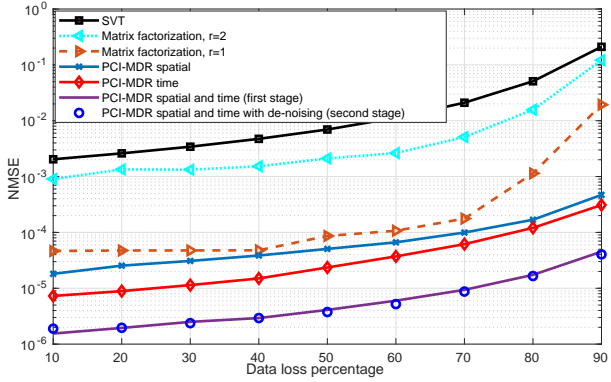}
 \caption{NMSE versus data loss for WSN data in the absence of noise}
 \label{fig:3}
 \end{figure}
 \begin{figure}[!ht]
  \centering
  \includegraphics[scale=0.6, trim=0 10 0 2]{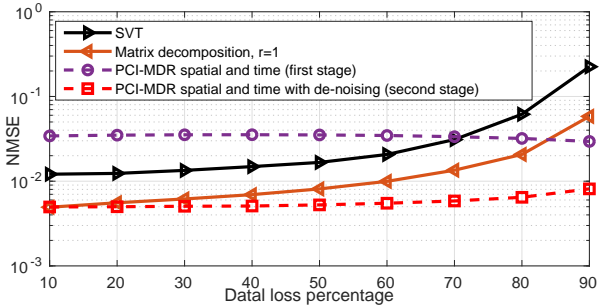}
  \caption{NMSE versus data loss for WSN data in the presence of noise}
  \label{fig:4}
   \end{figure}
Here, temperature dataset has 2699 missing entries from the total of 10600 entries ($53 \times 200$), i.e., $ 25.4623\%$ of data is not available. 

\textbf{Ground Truth:} Since the ground truth of the missing data mentioned above is not available, we manually removed $m_2$ entries randomly, in addition to the values already missing in the original dataset i.e., $m_1$. We designed the PCI matrix to recover the originally $m_1$ missing entries and also the simulated $m_2$ missing entries, therefore the overall missing data will be $N-M=m_1+m_2$.\\
\textbf{Data loss percentage} is given by $\frac{m_2}{N-m_1}\times 100 \%$.\\
\textbf{Normalized mean square error (NMSE)} is given as $\frac{||\textbf{x}_{m_2}-\hat{\textbf{x}}_{m_2}||_2^2}{||\textbf{x}_{m_2}||_2^2}$, where $\hat{\textbf{x}}_{m_2}$ is the recovered data at the simulated missing positions (i.e., $m_2$) and $\textbf{x}_{m_2}$ is the ground truth available at $m_2$ positions.

All the data points of \textbf{x} i.e., $N$ has been reconstructed, however, for computing the performance of the algorithm, the performance is calculated only at the simulated missing values, i.e., $m_2$ since the ground truth is available at these positions only. If the data at the simulated missing positions is recovered with good accuracy, it will indicate that the actual missing data must have also been recovered with good accuracy.

Fig. \ref{fig:3} and \ref{fig:4} shows the NMSE of data recovered at simulated missing positions as a function of data loss percentage in the absence and presence of noise, respectively. From Fig, \ref{fig:3}, we observe that \textit{a)} the reconstruction performance of the PCI-MDR method is better in the time domain as compared to the spatial domain. This is because the data in the time domain is smoother than the spatial domain and hence will be more sparse in the DCT domain, this is also evident from Fig. \ref{fig:1}; \textit{b)} PCI-MDR is performing better than both matrix factorization \cite{5961765} and SVT \cite{cai2010singular} methods. It may be noted that the matrix factorization technique also requires the rank of the data which is generally unavailable for the realistic scenario. Therefore, we have computed matrix factorization algorithm for two values of rank, $r$ = 1 and 2. Further, as the data loss percentage increases the performance of the proposed method compared to existing methods also increases.; \textit{c)} double sparsity can be exploited using both spatial and temporal domain together (\ref{12}) to further improve the performance of PCI-MDR.; and \textit{d)} at massive data loss of $90\%$, the proposed protocol is still giving NMSE of $4.0445 \times 10^{-5}$, however, the matrix factorization with $r=1$ has NMSE of 0.019348, and hence providing $ \sim 25$ dB improvement \footnote{The simulation codes for the results can be provided on request. }.


The second stage of de-noising in PCI-MDR method (\ref{13}) provides the significant improvement in the case of noisy data as shown in Fig. \ref{fig:4} but no improvement for noiseless data as shown in Fig. \ref{fig:3}. This is because in the presence of noise the smoothness of the real data gets affected, therefore exploiting low-rank constraint at the second stage provides an improvement in the reconstruction performance. However, in the absence of noise, the PCI and DCT based CS framework (first stage) is sufficient in recovering the missing data accurately, and hence no additional constraint such as low-rank is required. Compared to the existing matrix factorization algorithm with $r=1$, the proposed protocol provides 8.6 dB improvement in the presence of noise.

\section{Conclusion}
In this paper, a novel method, PCI-MDR has been proposed to recover missing data in wireless sensor networks. The method exploits the incoherency of PCI matrix with DCT in the CS based reconstruction for missing data recovery. It also exploits the low-rank behaviour of the data to de-noise the recovered data in the presence of noise. Simulation results on a real WSN dataset show that the proposed PCI-MDR algorithm outperforms various other well-known missing data recovery algorithms.

\bibliography{PCI-MDR}
\bibliographystyle{ieeetran}
\end{document}